\documentclass[conference]{IEEEtran}
\ifCLASSINFOpdf
  \usepackage[pdftex]{graphicx}
  \graphicspath{{../Ale/immagini/}{../jpeg/}}
  \DeclareGraphicsExtensions{.pdf,.jpeg,.png}
\else
\fi
\begin{document}
%
\title{Simulation tools to compare and optimize the mobility plans
}

\author{\IEEEauthorblockN{Alessandra Campo}
\IEEEauthorblockA{Dipartimento di Architettura\\
Universit\`a Roma Tre\\
Roma, Italy\\
Email: alessandracampo.88@gmail.com}
\and
\IEEEauthorblockN{Roberto D'Autilia}
\IEEEauthorblockA{Dipartimento di Architettura\\
Universit\`a Roma Tre\\
Roma, Italy\\
Email: roberto.dautilia@uniroma3.it}
}


%


\maketitle

\begin{abstract}
	In the last decades, mobility planning has been a fundamental issue for the development of cities. A full knowledge of the way a mobility system influences the traffic behavior of a whole city is needed in order to propose plans aligned with the municipalities' goals. In particular, a tool to compare different plans and the respective costs and benefits is necessary to forecast the consequences of the changes in mobility plans.
	The aim of this research is to show how two different mobility models can be compared, based on different plans, and how to use this comparison strategy to develop or review a mobility plan.
	In fact, the results of a model simulation provide a detailed analysis of the use of every infrastructure, giving the municipality a tool to evaluate mobility plans and to choose the best solution.
	We considered samples of different urban mobility plans of the same city to show how the global behavior changes while changing infrastructures.
	As a case study we analyzed the city of Barcelona, where in the last 10 years the municipality applied a new urban mobility plan, reviewed every 5 years (PMU 2007 and 2012). By using MATSim simulation tools, we realized two models of the infrastructure network: in both we built metro, train, tram and car networks. The main differences between the two models are the speed limit of the car network and the presence or absence of the bicycle network.
	The results show that if we set low speed limits and increase bike infrastructures, the average travel time decreases and the number of users of the bicycle network increases to the detriment of public and private transport. In particular, we show that the number of users of public transport decreases more than the private transport users. This behavior is related to the lack of the bus network, not simulated in these models, suggesting that the metro network works properly only when it's integrated with a bus network.
\end{abstract}


%
\IEEEpeerreviewmaketitle

\section{Introduction}
During the 19th century the cities changed radically due to the fast demographic and economic growth, caused in turn by the industrialization and the immigration from the countryside.
After the Second World War, the western world reached a newfound wealth that, combined with the technological development and easy access to cars, led the cities to expand at the cost of congested streets, air pollution and noise \cite{falcocchio2015}.
In order to tackle these issues, reorganize the cities and slow the private transport trend, municipalities developed mobility plans.

In this scenario, it is necessary to introduce methodologies to evaluate plans and verify that the consequences of the changes in the mobility plans are aligned with the municipality's goals.The urgency to find a realistic method to analyze plans is due to the rapid growth of the city dimensions and its complexity, features very hard to manage without suitable computation tools \cite{angel2012atlas}.

The difficulty to predict the evolution of complex systems, such as mobility, is due to the collective behavior of many conflicting variables, where the examination of one variable at a time leads to wrong conclusions \cite{citeulike:1398137}.
This suggests the necessity of using simulation models for urban planning and computational tools to predict the emerging behaviors.

Breakthroughs for mobility planning are represented by the Wardrop \cite{wardrop1952} and Braess \cite{Braess1968} results.
Wardrop developed two principles, related to the Nash Equilibrium \cite{nash1950}\cite{Nash1951}, defining the concept of user equilibriums.
Braess, instead, found an important paradox in traffic flow, showing that even for a simple 4-roads network, the overall journey time can increase when adding a new route \cite{Colak2016}. 
These results make evident that the optimization of large traffic network is not a trivial task.

Thanks to increasing computer speed it has been possible to test these ideas on large synthetic networks.
The simulation of agent-based models depends on computer speed; however, at the moment, it is the only way to extract data for a given scenario.
In this framework several models have been developed and different real cases considered.
The effects of a new airport in Berlin, a possible motorway in Zurich, the emissions in the Munich city, the evacuation after a chemical accident in Aliaga or the impact of bike sharing in Barcelona \cite{matsim2016} are only a few examples.

Having a major effect on the quality of people's lives, mobility plans are an important section of city policy.
Without the help of theoretical models, however, the complexity of cities makes mobility management very challenging.
As evidenced by the Braess paradox, it is difficult to forecast optimal traffic strategies even for small networks with simple agents' plans, raising the need of manageable models to evaluate the potential of mobility plans in advance.

In this paper we use the MATSim microsimulation tools for the Barcelona case study showing how to test the effects of a mobility plan before it is implemented and to extract indications for possible criticalities or improvements.
For this purpose two different scenarios of Barcelona Mobility Plan (PMU) have been created: the first has been designed according to the mobility plan of 2007, while the other is based on the current situation.
The results show that the real effects of the PMUs in the city are similar to the effects we obtained through simulations and that they are congruent with the municipality's intentions.

\section{The MATSim Microsimulation}
In 1968 Braess analyzed the Nash Equilibrium for two simple road systems, measuring the users' travel time.
The two systems were a 4-road and a 5-road network with a given travel time for each road, eventually depending on the number of cars using that road \cite{Colak2016}. 
Comparing the overall travel time for the two networks at the Nash Equilibrium, he found that the removal of a road could actually improve the users' travel time.

The simulation of the same networks with MATSim \cite{matsim2016} shows that, in general, the system does not reach the Nash Equilibrium.
If 100 cars are left evolving toward optimal strategies on the two networks, the system in general does not reach the average travel time given by the Nash Equilibrium.
In fact the MATSim Montecarlo algorithm does not try to reach the Nash Equilibrium, but rather a Stochastic User Equilibrium, where the {\sl traveller selects the route which he or she perceives to have minimum cost} \cite{Hazelton1998101}.

\begin{figure}[!t]
\centering
\includegraphics[width=3.0in]{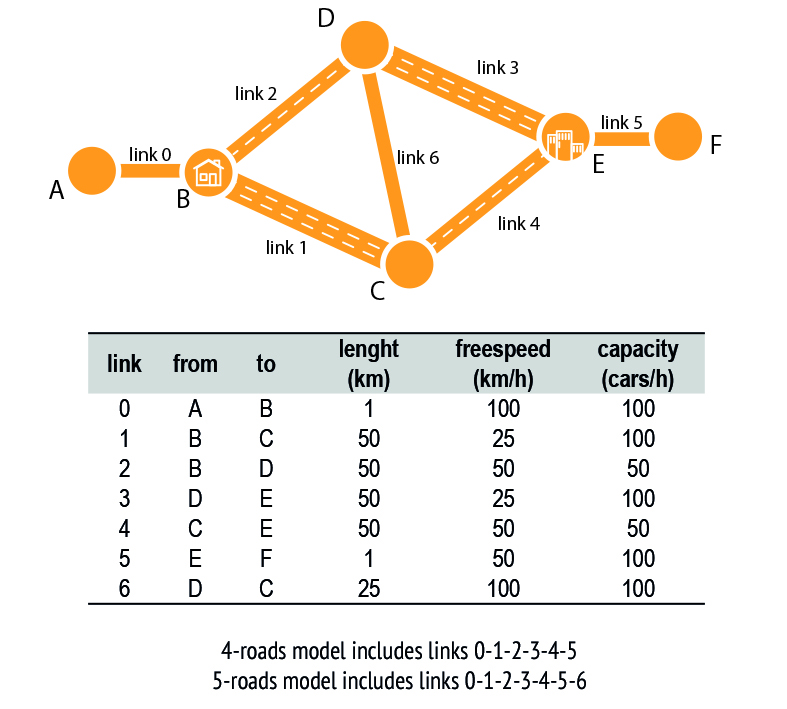}
\caption{Braess Paradox simulated with MATSim}
\label{fig0_BraessParadox}
\end{figure}

We set the parameters of capacity, length and speed limit in order to have the same travel time in the different streets as the theoretical example had set, as it's shown in Fig.\ref{fig0_BraessParadox} 

In both the models (4-road and 5-road) it's possible to see a clear difference with the Nash Equilibrium.
In the 4-road model cars don't split equally in the two possible routes, instead 33\% choose the route on links 2-3, and the other 67\% choose links 1-4.
In this way all cars using  2-3 links reach their goal in 3h (1h+2h), while the first car on the route 1-4 reaches node F in 3h (2h+1h) and the last one in 3h20m (2h+1h20m), due to the capacity of link 4 set to 50 cars/h, which causes a slowdown.

The same simulation applied to the 5-road network clearly shows how the route choices are influenced by the flow of time.
In fact, at the beginning, the cars have to choose between link 1 or 2 based on their convenience and they arrived in nodes 2 and 3 at different times.
Once they reach central nodes they instantly decide which link they are going to use, without waiting for the other cars. 

This behavior brings the following cars to choose their own route based on the current situation, differing from the theoretical example of Braess Paradox.
For this reason all cars choose the route 2-6-4 to reach node F, ignoring the link 1 and 3.
In fact cars don't have any advantage to choose them because travel time is set at 2 hours, when the travel time of link 2 and 4 can be between 1 and 2 hours, depending on the cars that are using that link.
So the first car that enters in the link 2 needs 1 hour to reach node D, instead, the last car needs 2 hours for the same route, due to the capacity of the street. This difference brings the first car reaching node D to use the link 6 and 4, which are still empty, and reaches its goal in 2h15m (route 2-6-4 = 1h+15m+1h), while the last car needs 3h15m (route 2-6-4 = 2h+15m+1h).
Through these simulations it's possible to notice that the equilibrium reached by the MATSim Montecarlo is influenced by time, unlike the Nash Equilibrium, where everyone chooses their strategy at the same time.

\section{The Barcelona Network}
To evaluate and compare the 2007 and 2016 Barcelona Mobility Plans \cite{PlaMobilitat20132018}\cite{PlaMobilitat2008}, two MATSim models of the city have been created, taking into account some variables as population, activity, cycling and car networks, and public transportation.
The main differences between the two models are the presence of the bicycle network, the speed limit and three new metro lines in the 2016 model.

The area taken into account includes the whole transportation system of Barcelona and the closest municipalities ({\sl L'Hospitalet de Llobregat}, {\sl El Prat de Llobregat}, {\sl Cornell\`a de Llobregat} and {\sl Badalona}).
Considering the impossibility of simulating the entire population of the area, a synthetic one was generated based on data of the Department of Statistics of the Municipality of Barcelona \cite{BarcelonaEstadistica}, and a representative sample of 10\% of the total of 1.6 million of people has been developed. 

In order to build a detailed population mobility demand, several statistics have been examined: occupations, activities distribution and location, average distance between the start and the end of agent's plan and modes of transport.
All this information, based on the Department of Statistics data \cite{BarcelonaEstadistica}, has been combined to define the activities and the synthetic population of the model.

Four main activities categories were created with a total of 9 types: home, work, education (school, high school, university), recreation (shopping, sport, leisure, restaurant).
For each activity the location and the hours of opening were defined.
The location of home, high school, leisure, restaurant and work were set according to a bivariate Gaussian distribution with mean at the center of the city.
The school and sport locations were supposed to be respectively inside a radius of 2 and 3 km from the home.
Some shops were also set according to a bivariate Gaussian distribution with mean in the center of the city and malls and commercial streets have been placed in the current position.

Two different network infrastructures have been built.
In the 2007 model, the city center is almost entirely a pedestrian area, while the other streets have the 50 km/h speed limit.
The 2007 model does not consider bikes as a relevant vehicle because in 2007, although bicycles were allowed everywhere, only few bicycle lanes were built.
Therefore, for simplicity, we did not consider the 2007 bicycle network.

For the 2016 model, according to the PMU purpose, two types of roads were defined.
The first one, that includes main roads, kept the 50 km/h speed limit while the other ones set the limit at 30 km/h.
In this scenario, due to the large investment of the municipality on the cycle paths and sustainable mobility, the bicycle infrastructure has been introduced.

The city trains, the tram and the metro lines were modeled into the infrastructure network, but the bus lines were omitted.
This is a serious limitation of the model, but it is still useful to estimate a lower bound for the overall network behavior.
The main difference between the two models is given by the introduction of three new metro lines (Line 9 north, Line 9 south, and Line 10) in the 2016 model.
The other differences between 2007 and 2016 infrastructures were not considered because they involved little changes that were not significant for the purposes of this analysis.

\section{Results}
\begin{figure}[!t]
\centering
\includegraphics[width=3.0in]{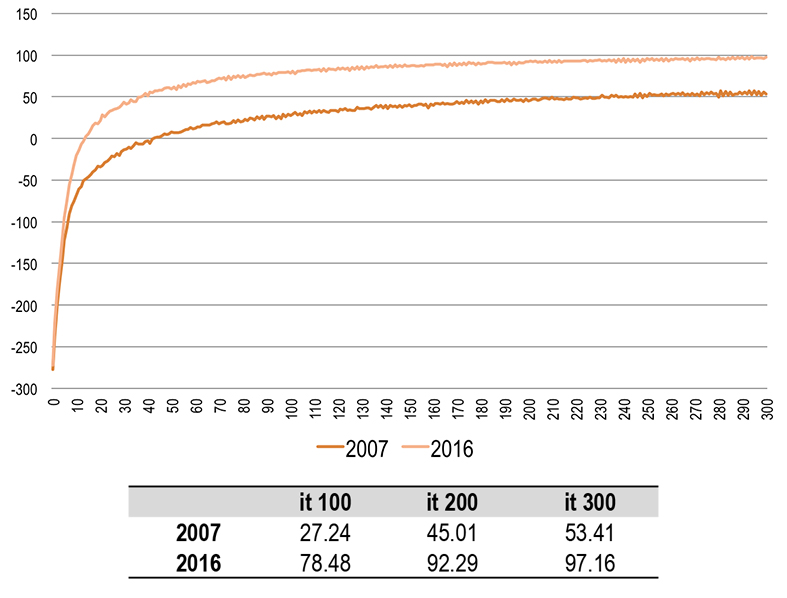}
\caption{Comparison between average executed score of the 2007 and the 2016 models}
\label{fig1_average_executed_score}
\end{figure}

\begin{figure}[!t]
\centering
\includegraphics[width=3.0in]{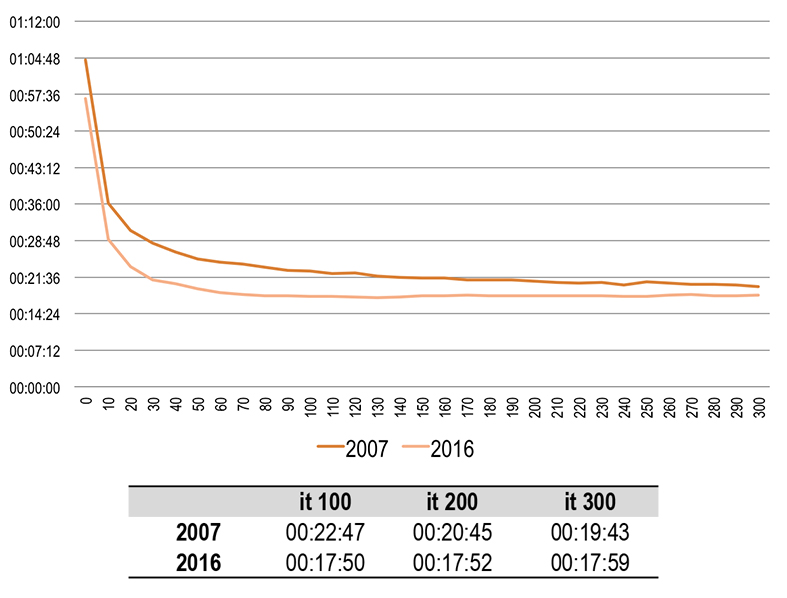}
\caption{Comparison between average trip duration of the 2007 and the 2016 models}
\label{fig2_average_trip_duration}
\end{figure}

The results of the computer simulations show the different response of the two transport networks to the travel demand.
To reach the Stochastic User Equilibrium the Montecarlo simulation has been iterated 300 times, the convergence is shown in Fig.\ref{fig1_average_executed_score}.
We observe that the score of the 2016 scenario is higher than the 2007 one.
On average, in fact, for the 2016 scenario, every agent found better routes or faster means of transportation.
This is confirmed by the comparison of the average trip durations in Fig.\ref{fig2_average_trip_duration} where an improvement of almost 2 minutes is observed for the 2016 network with respect to the 2007 one.

\begin{figure}[!t]
\centering
\includegraphics[width=3.0in]{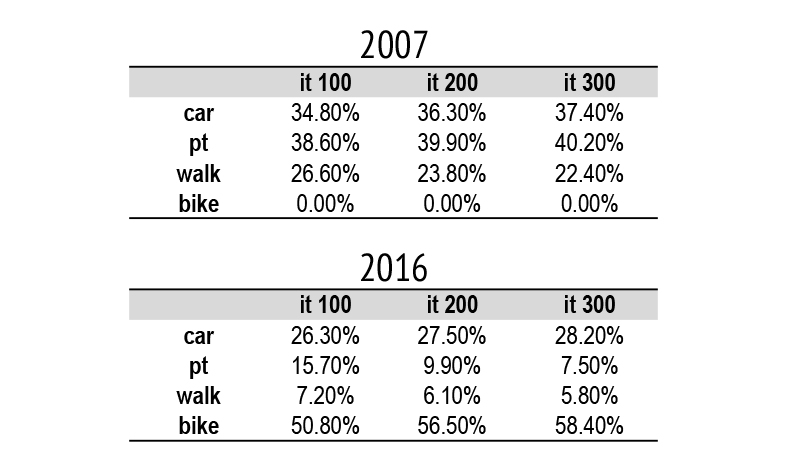}
\caption{Means of transportation in the 2007 and the 2016 models}
\label{fig4_means_of_transportation}
\end{figure}

We also computed the percentage of use of car, public transport, walk, bike at different iteration stages.
In Fig.\ref{fig4_means_of_transportation}, it's possible to see the percentage of usage for the different means of transport for the two scenarios, at the iteration 100, 200 and 300.

\begin{figure}[!t]
\centering
\includegraphics[width=3.0in]{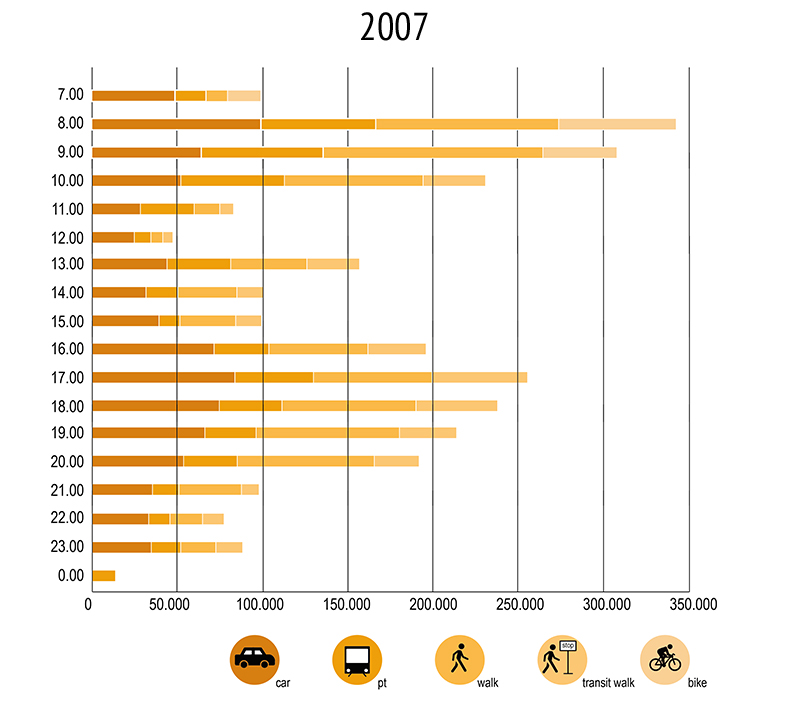}
\caption{Traveling people in the 2007 model}
\label{fig5_traveling_people_2007}
\end{figure}

\begin{figure}[!t]
\centering
\includegraphics[width=3.0in]{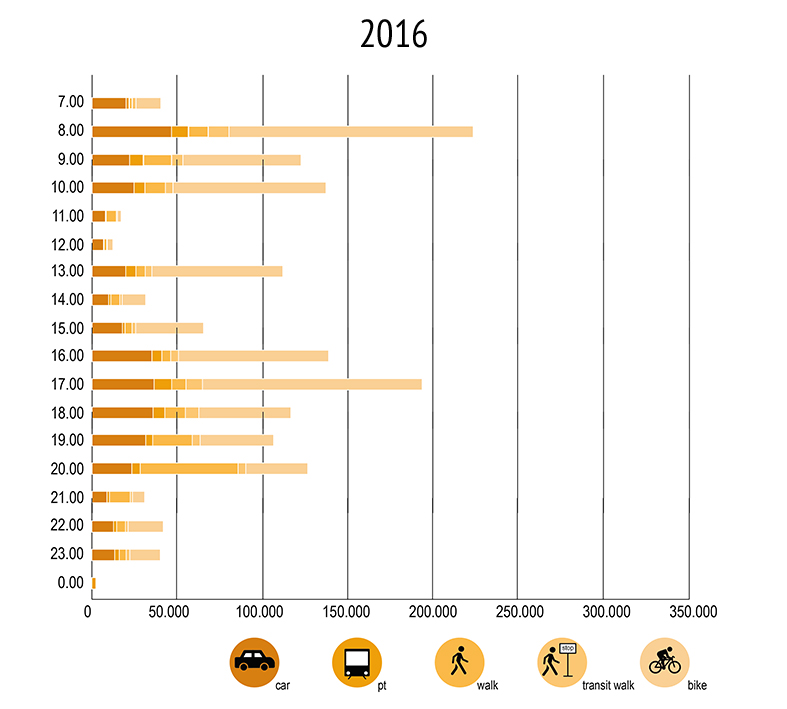}
\caption{Traveling people in the 2016 model}
\label{fig6_traveling_people_2016}
\end{figure}

The 2007 simulation results show the use of means is almost defined at the iteration 100, but the average trip duration keeps improving from 00:22:47 to 00:19:43.
This apparent incongruence demonstrates that agents don't change means to improve their time travel, but rather they reduce the travel time by changing their routes.
The 2016 data, on the contrary, show that public transport and bike percentages change from iteration 100 to iteration 300 without a relevant improvement of the average time travel.
The comparison of traveling people on the two scenarios, Fig.\ref{fig5_traveling_people_2007},\ref{fig6_traveling_people_2016}, shows that, due to the reduced travel time,  fewer people are moving in the city.

\begin{figure}[!t]
\centering
\includegraphics[width=3.0in]{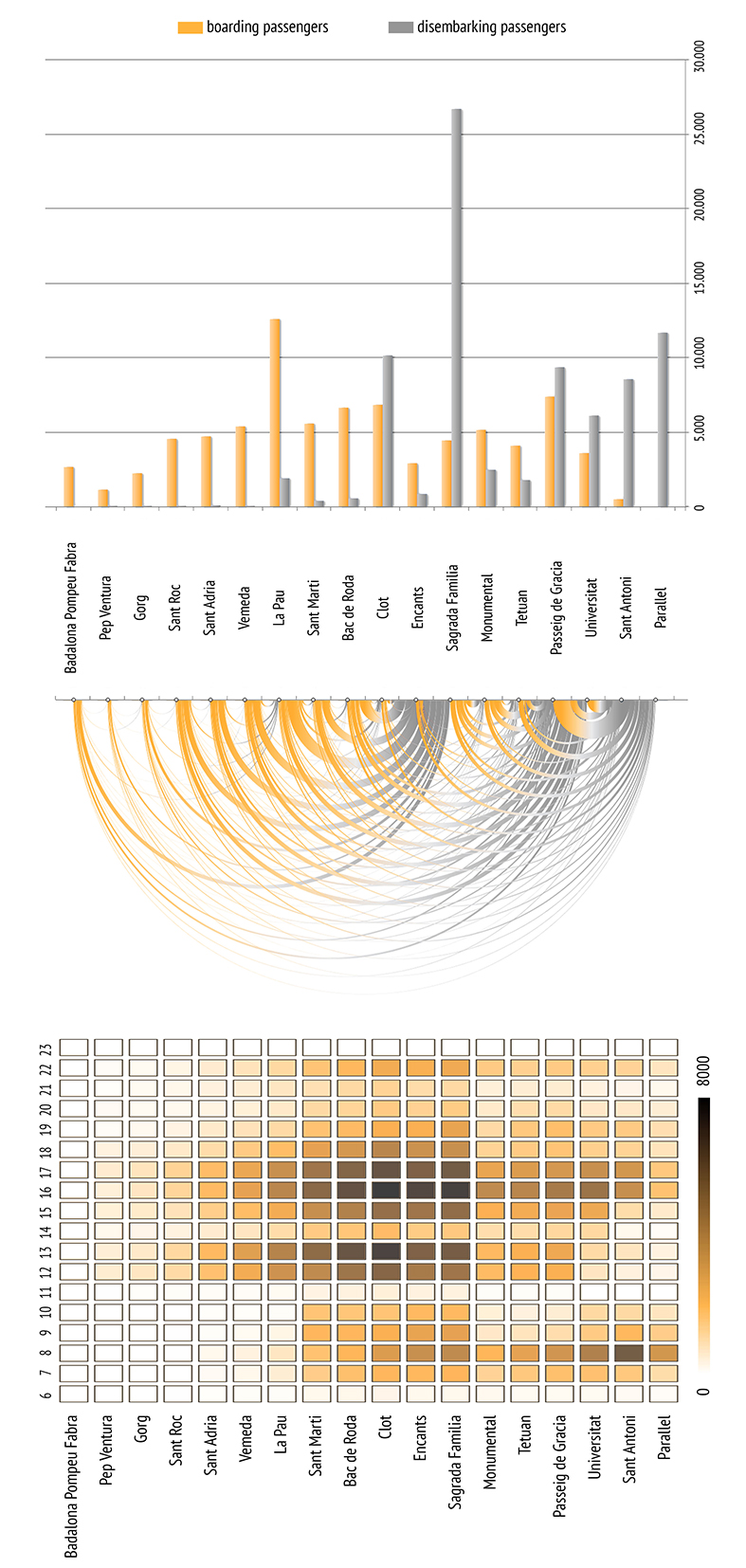}
\caption{Analysis of line L2 from Badalona Pompeu Fabra to Parallel in the 2016 model}
\label{fig7_L2_BadalonaParallel}
\end{figure}

The output data make it possible to analyze the metro system in detail.
If we consider, for example, data from the metro line L2 in the 2007 scenario.
The line connects the northeast of the city with the city center, and it is 13,1 km long.
Fig.\ref{fig7_L2_BadalonaParallel} shows that, during the morning, the trains directed to the suburbs are mostly used between the stops {\sl Passeig de Gracia} and {\sl La Pau}, and the main stations where people get on board are {\sl Parallel}, {\sl Universitat} and {\sl Sagrada Familia}, but there isn't a particular station where people get off the train.

\begin{figure}[!t]
\centering
\includegraphics[width=3.0in]{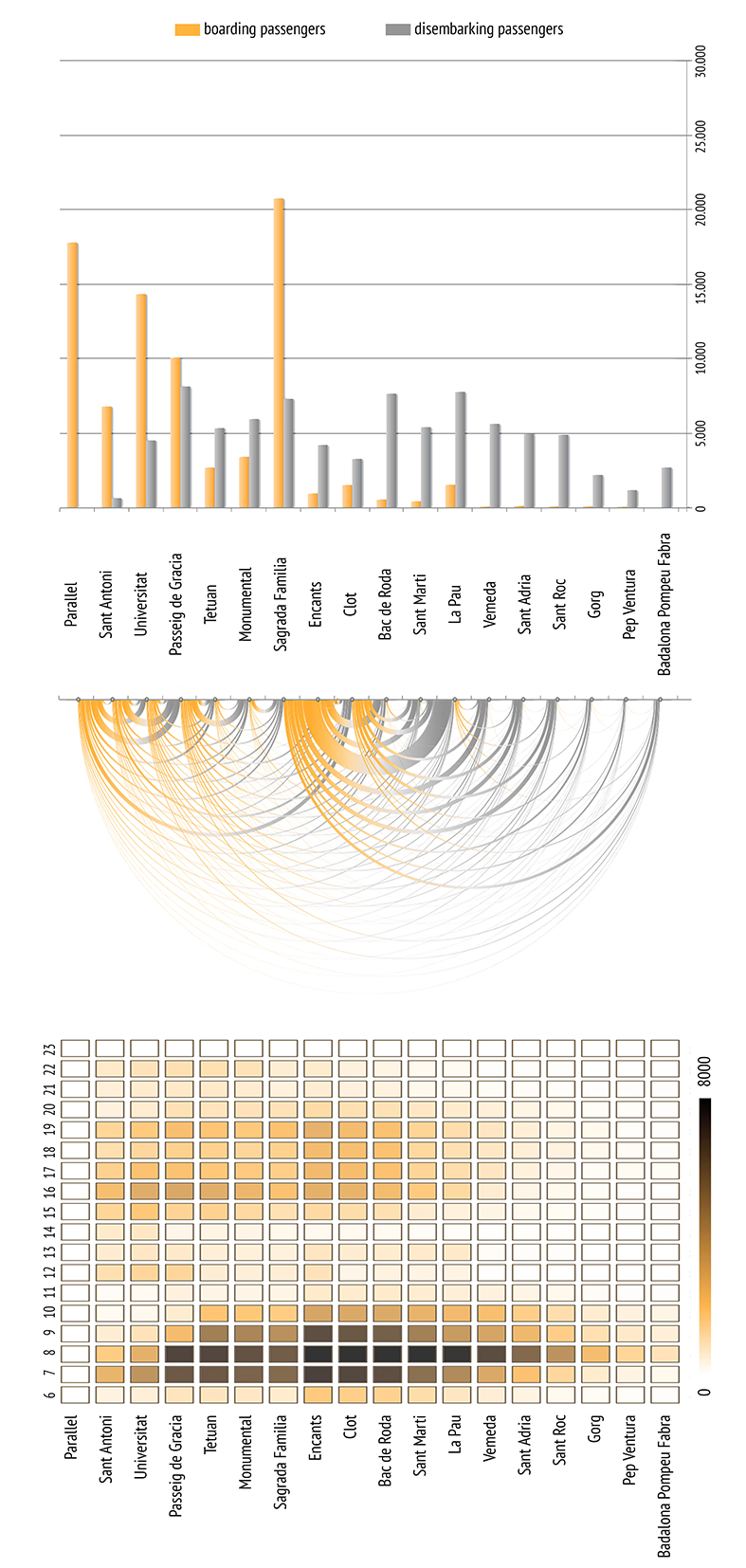}
\caption{Analysis of line L2 from Parallel to Badalona Pompeu Fabra in the 2016 model}
\label{fig8_L2_ParallelBadalona}
\end{figure}

If the opposite direction is taken into consideration, it's possible to notice that the trains reach almost complete saturation between the stops {\sl Sant Marti} and {\sl Sagrada Familia}, from 4:00 pm to 6:00 pm.
The main station where people get on board is {\sl La Pau}, while the others have approximately the same flow.
Fig.\ref{fig8_L2_ParallelBadalona} shows that {\sl Sagrada Familia} is where the majority of people get off board, followed by {\sl Parallel}, {\sl Sant Antoni} and {\sl Passeig de Gracia}.

\begin{figure}[!t]
\centering
\includegraphics[width=3.0in]{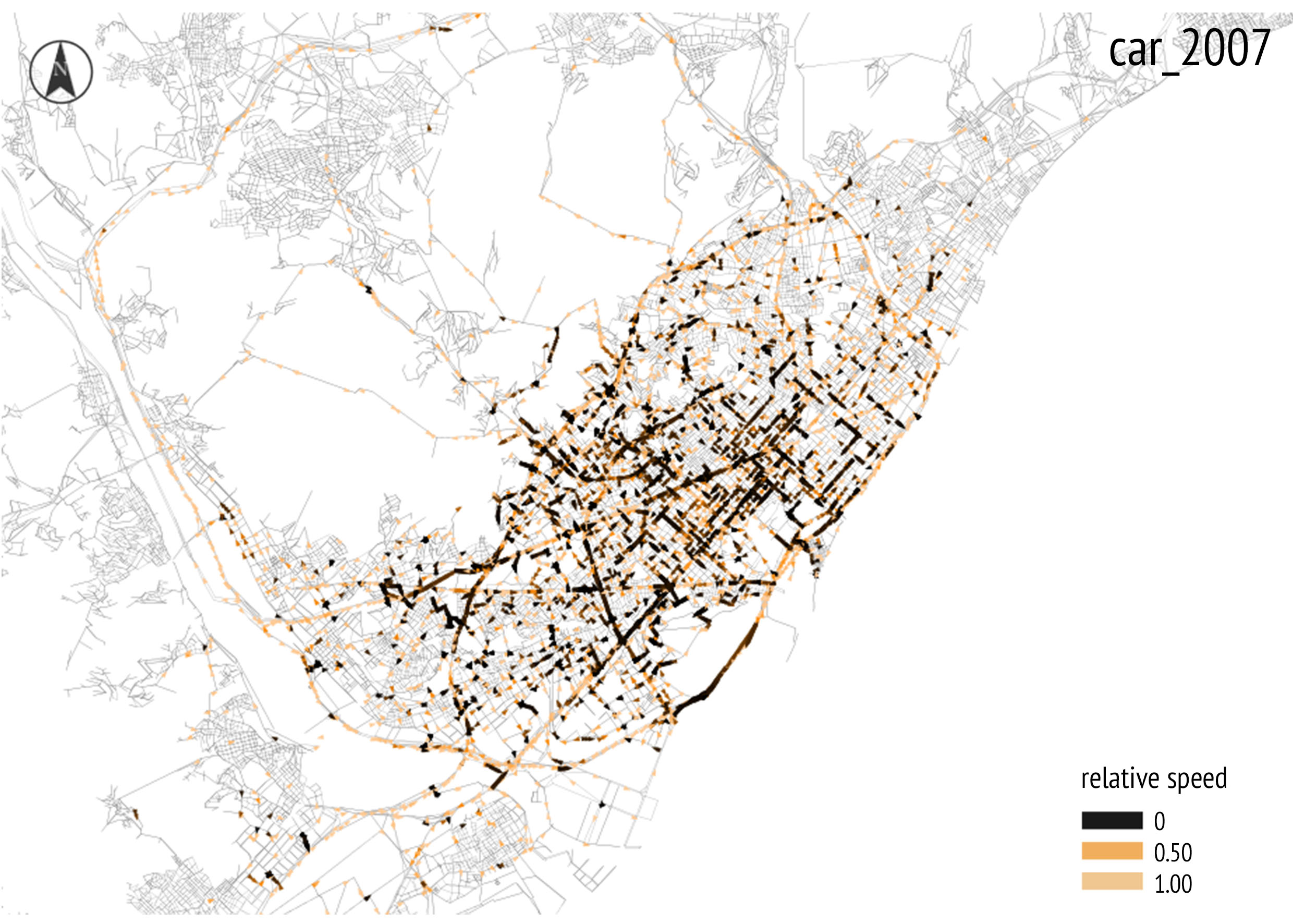}
\caption{Relative car speed in the 2007 model}
\label{fig9a_car2007}
\end{figure}

\begin{figure}[!t]
\centering
\includegraphics[width=3.0in]{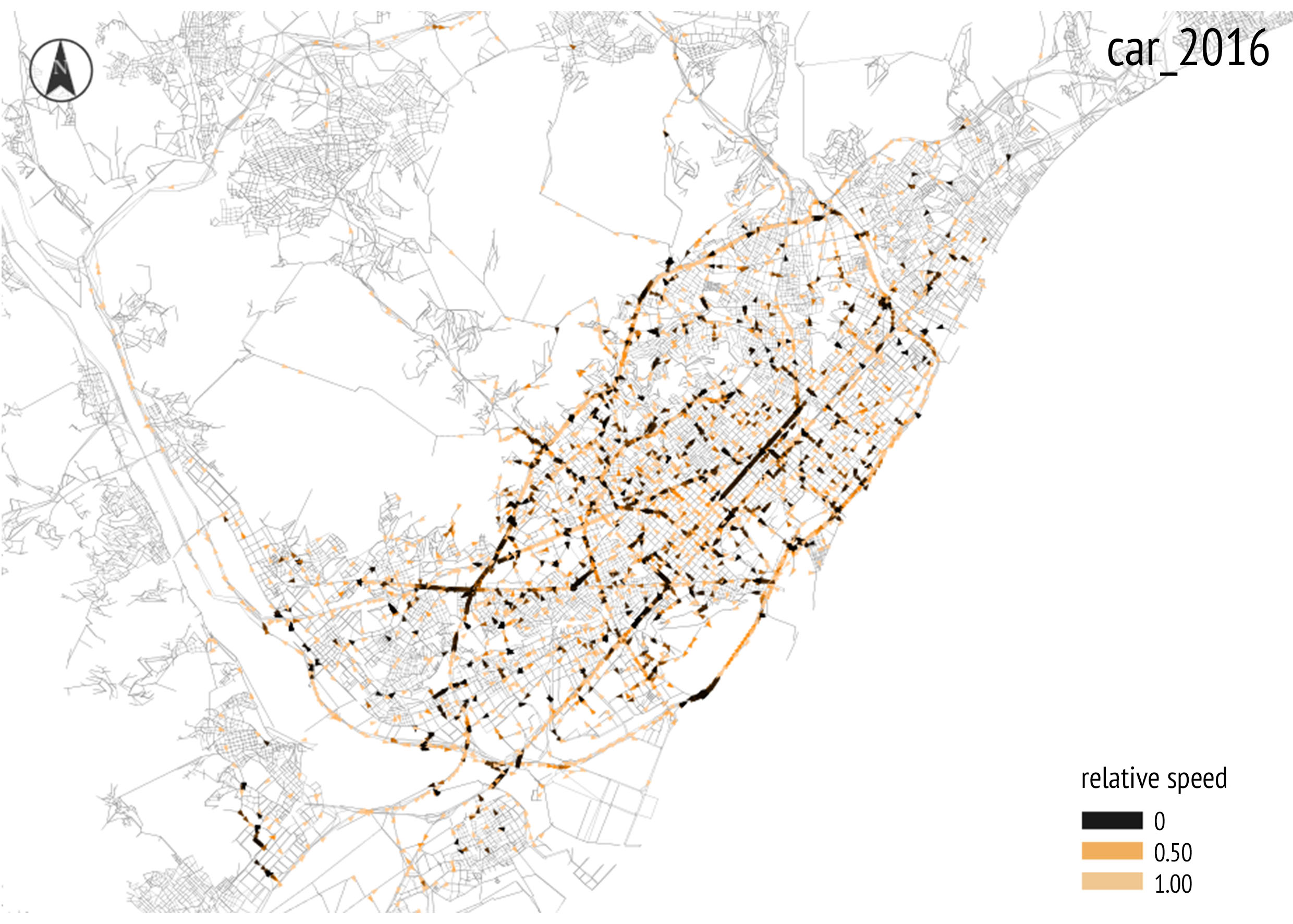}
\caption{Relative car speed in the 2016 model}
\label{fig9b_car2016}
\end{figure}

\begin{figure}[!t]
\centering
\includegraphics[width=3.0in]{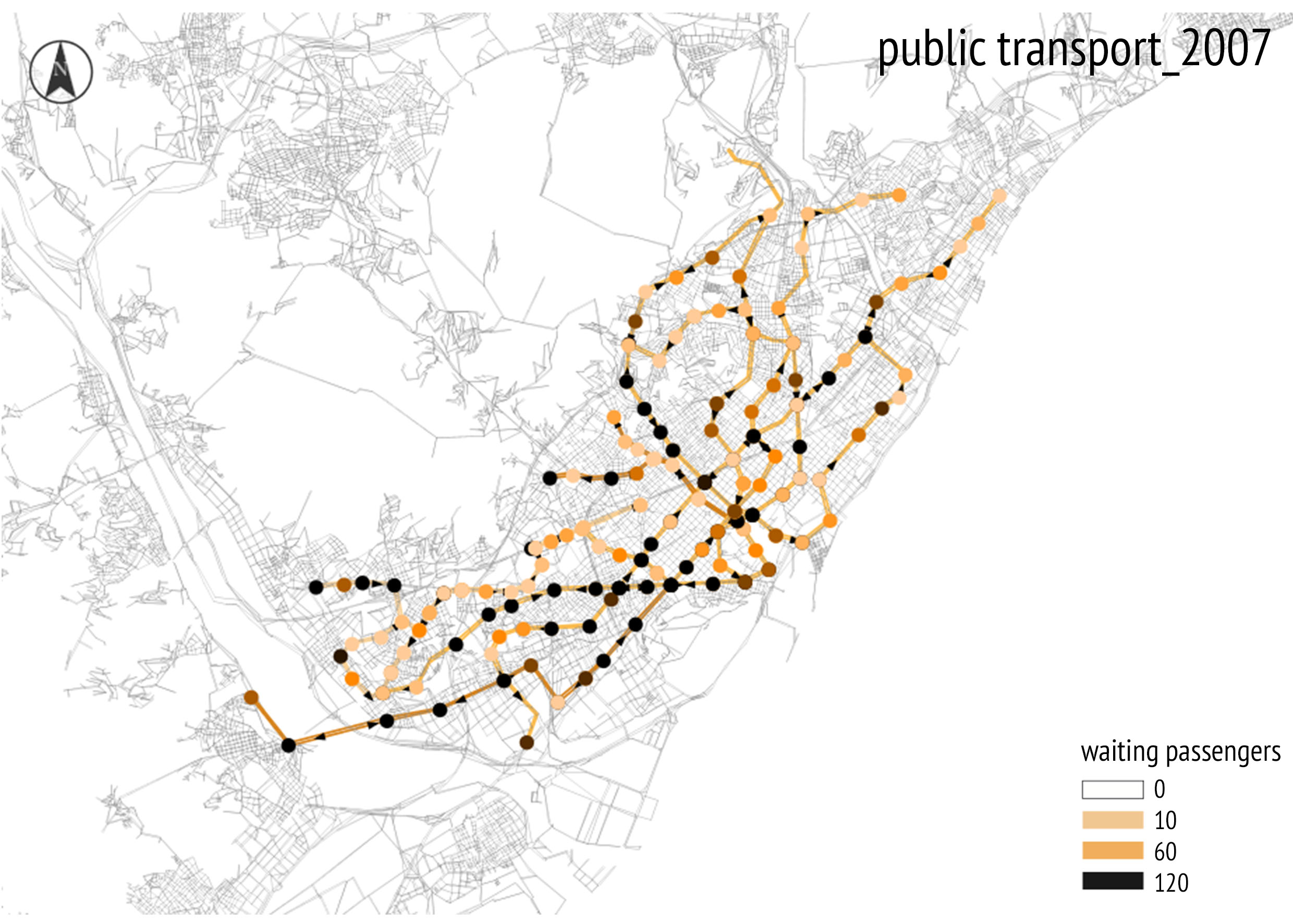}
\caption{Saturation of public transport stops in the 2007 model}
\label{fig10a_pt2007}
\end{figure}

\begin{figure}[!t]
\centering
\includegraphics[width=3.0in]{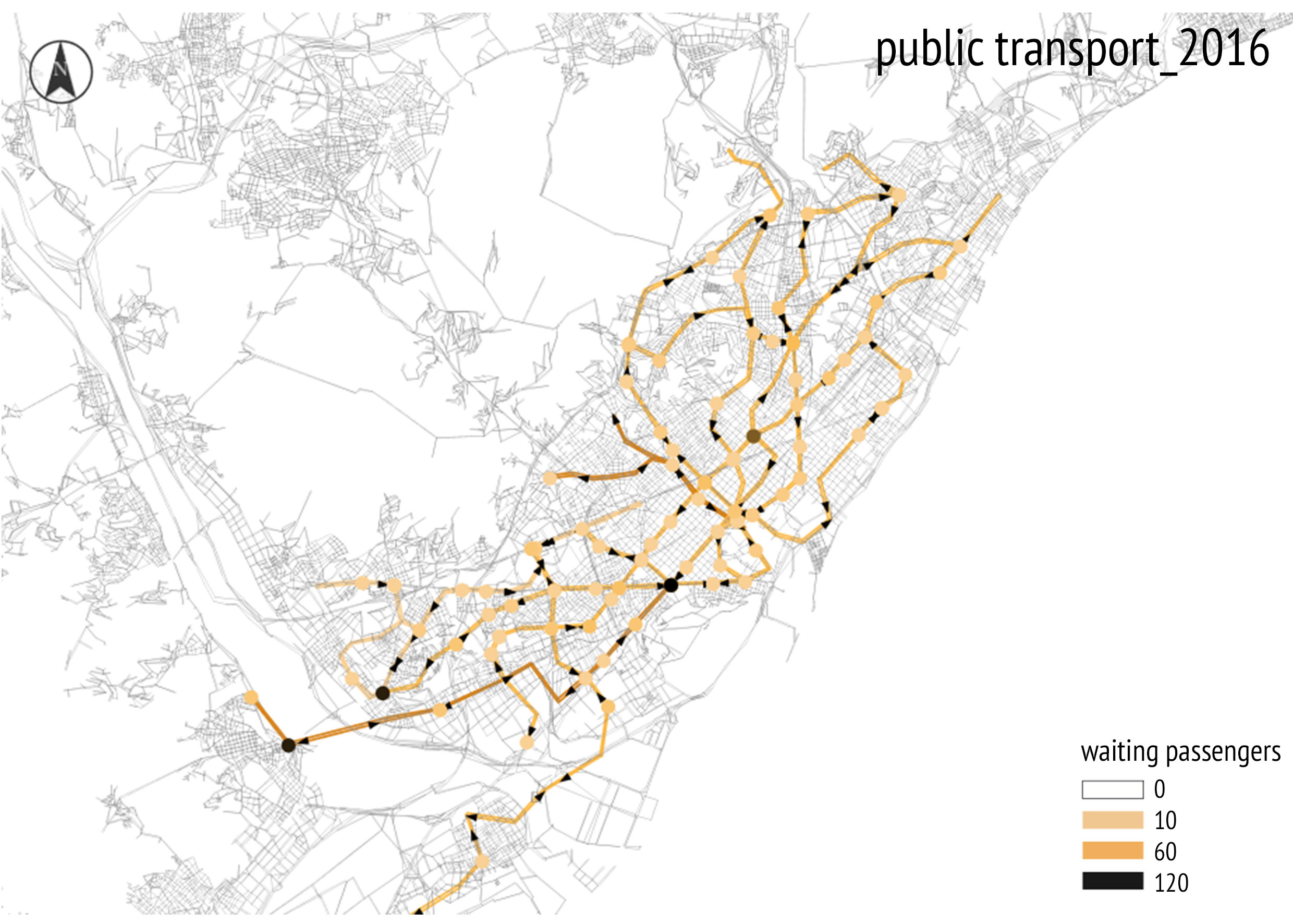}
\caption{Saturation of public transport stops in the 2016 model}
\label{fig10b_pt2016}
\end{figure}

The overall dynamical view of the simulation results shows the traffic jams and the infrastructure's saturation for the two scenarios, Fig.\ref{fig9a_car2007},\ref{fig9b_car2016},\ref{fig10a_pt2007},\ref{fig10b_pt2016}.

\section{Conclusion }
This work shows how the data of the simulation models can be used to compare and develop mobility plans. 
However, it is important to stress that the method described is effective when used to evaluate mobility plans which have already been created.
In fact, urban plans are a compromise among numerous factors, as political mediation, economic impact, time issues etc.
Therefore, it is necessary to develop them taking all these factors into consideration, not on basis of statistical models or multi-agent simulations.
In turn, simulation methods are extremely useful for {\sl a posteriori} evaluation of these plans.

The analysis of the two Barcelona mobility plans shows an improvement from 2007 to 2016 in terms of travel time and network congestion.
This result is confirmed by the decrease of the average trip duration together with the decrease of the use of cars and public transportation in favor of more sustainable means, such as bikes.
This behavior, resulted in the simulation, is aligned with the Barcelona municipality's goals.
The purpose of the plan was, in fact, to overturn the use of means, prioritizing walking and the use of bikes, and then public transport, taxi and cars.

Also, analyzing and comparing the two models, in the 2016 scenario it is possible to see an increase of traffic flow on the streets with a 50 km/h speed limit and a decrease of traffic flow on the streets with a speed limit reduced to 30 km/h.
Again, this change on the route choice coincides with the municipality's goals, which planned the use of the minor streets only for residents.

The obvious decrease on the use of public transport, from the 2007 model to the 2016 model, indicates that the absence of the bus network influences strongly the entire system. Nowadays, there are no cities in the world that have a metro network without a bus network.
In fact, the range of influence of a metro station is approximately 1 km, and, without a bus infrastructure, the areas situated outside this range, wouldn't benefit from the metro infrastructure, which would lead to preferring other means, mainly in the short distance trips.

In conclusion, these analyses show that, through the comparison of the two models, we obtained results comparable with the current real situation, and that, if the 2016 model had been built during the development of the mobility plan, the municipality would have known the consequences of this plan in advance.
\section*{Acknowledgment}
The authors would like to thank Mario Cerasoli, Simone Ombuen and Camilla Desideri for their valuable comments and observations.



%




\end{document}